\documentclass[aps,pre,onecolumn,eps,graphicx]{revtex4}
\def\bacol{\setlength{\arraycolsep}{0pt}}
\def\bec{\begin{center}}
\def\enc{\end{center}}
\def\be{\[}
\def\ben{\begin{equation}}
\def\ba{\begin{array}}
\def\bea{\begin{eqnarray}}
\def\ee{\]}
\def\een{\end{equation}}
\def\eea{\end{eqnarray}}
\def\ea{\end{array}}
\def\btab{\begin{table}}
\def\btabu{\begin{tabular}}
\def\etab{\end{table}}
\def\etabu{\end{tabular}}
\def\bit{\begin{itemize}}
\def\eit{\end{itemize}}
\def\bef{\begin{figure}[htb]}
\def\befh{\begin{figure}[!h!]}
\def\enf{\end{figure}}

\def\a{\alpha}

\def\gb{\beta}

\def\e{\epsilon}
\def\g{\gamma}
\def\barg{\bar{\gamma}}
\def\bara{\bar{\alpha}}

\def\l{\lambda}

\def\s{\sigma}

\def\half{{\textstyle{1 \over 2}}}

\def\b1{{\bf 1}}

\def\bp{\mbox{\boldmath $p$}}

\def\cos{\hbox{cos}\:}

\def\cosh{\hbox{cosh}}

\def\nn{\nonumber}
\def\bb{\left(}
\def\eb{\right)}

\usepackage{epsfig,latexsym}

\newcommand \bew {\begin{widetext}}
\newcommand \enw {\end{widetext}}
\newcommand \eps {\epsilon}

\begin{document}

\title{\bf\noindent Electrostatic Fluctuations in Soap Films}

\author{D.S. Dean$^{(1)}$ and R.R. Horgan$^{(2)}$}

\affiliation{(1) IRSAMC, Laboratoire de Physique Quantique, Universit\'e Paul Sabatier, 118 route de Narbonne, 31062 Toulouse Cedex 04, France\\
(2) DAMTP, CMS, University of Cambridge, Cambridge, CB3 0WA, UK \\
E-Mail:dean@irsamc.ups-tlse.fr, rrh@damtp.cam.ac.uk}

\begin{abstract}
A field theory to describe electrostatic interactions in soap
films,  described by  electric multi-layers with a 
generalized thermodynamic surface-charging mechanism,
is studied. In the limit where the electrostatic interactions
are weak this theory is exactly soluble. The theory incorporates
in a consistent way, the surface-charging mechanism and the 
fluctuations in the electrostatic field which correspond to the
zero frequency component of the van der Waals force. It is 
shown that these terms lead to a Casimir like attraction 
which can be sufficiently large to explain the transition
between the Common Black Film to a Newton Black Film. 
\end{abstract}
\maketitle
\vspace{.2cm}
\pagenumbering{arabic} 
\pagestyle{plain}

\section{Introduction}
Soap films are naturally occurring examples of diffuse double layers
\cite{is,rusasc}. They
consist of two surfaces formed by hydrophobic surfactants which 
accumulate at the air water interface due to the hydrophobic nature
of the hydrocarbon chains of the surfactant molecule. In the classic 
experimental set up to measure the disjoining pressure in 
soap films, the film is placed in a cell in contact with a bulk
solution, usually in a porous frit
\cite{my,exkokh,dese,caetal}. A capillary connecting the bulk
solution to the outside of the  cell allows the pressure in the cell 
to be varied and a direct measurement of the disjoining pressure $P_d$,
generated by interactions within the film, is thus possible. 
The thickness of the film as 
a function of  $P_d$, along with certain aspects of its structure such as
the electronic density, can then be measured via X-ray or optical methods.
The interface formed by the surfactant can lead to the formation of a surface 
charge and, in addition to the surfactant, the bulk can contain an electrolyte
such as salt. 

The simplest model of a soap film is of two charged surfaces separated by
an electrolytic solution. Electrostatic interactions therefore play an
fundamental role and can be taken into account at the simplest level of 
approximation by the mean field Poisson
Boltzmann equation \cite{is,rusasc}. 
In the simplest models, the surface charge or surface 
potential is taken to be fixed and is hence a fitting parameter of the 
theory. More elaborate theories take into account the mechanism of surface 
charging \cite{nipa1,dese,pisa}. 
Steric effects between the counter-ions in the vicinity of the 
surface, known as the Stern layer, may also be incorporated
\cite{orsteric}. In general
the disjoining pressure predicted by such theories is repulsive. In 
a detailed experimental study of soap films formed by ionic surfactants
such as sodium dodecyl sulphate (SDS) 
\cite{dese}, it was shown that a modified Poisson Boltzmann theory 
incorporating an energetic surface-charging mechanism, via an 
attractive free energy potential at the film surface, can predict 
the surface tension of the bulk solution. Then, with no free fitting 
parameters remaining, the theory predicts the disjoining pressure of the 
films made from this bulk, just up to a pressure where the Common Black Film
(CBF) collapses and forms a Newton Black Film (NBF). The NBF is an extremely
thin film where experiments \cite{bebe}
show that there is an extremely small separation
between the two surfactant surfaces (about one layer of water molecules 
across). The transition between the CBF-NBF is experimentally interpreted
as a first order phase transition \cite{caetal}, and the  body of 
theoretical work supports this interpretation. The rough picture, coming from
the approximate theories that exist, is that at large distances the
disjoining pressure is repulsive and stabilizes the CBF. At closer 
inter-surface distances attractive van der Waals forces some into play 
which are responsible for the eventual collapse to the NBF. 

The aim of this paper is to analyze
the role of electrostatic fluctuations in the presence  of generic 
surface-charging mechanisms and spatially varying dielectric constants 
in diffuse layer systems. Previous studies incorporating the fluctuations 
about the mean field Poisson Boltzmann theory in simpler models have revealed
that these fluctuations lead to attractive interactions 
\cite{po,codu,btco}. 
The analysis here is adapted to the limit of weak electrostatic
interactions. We use a field theoretic formulation of the problem, which
in addition to producing new results for fluctuating surface charges, 
allows one to recover, in a powerful and universal way, many results already
established in the literature. In addition  our formulation is well
adapted  to develop a perturbative expansion which allows one to incorporate
higher order interactions.

The Poisson Boltzmann 
theory may be supplemented by including the 
contribution of dispersion forces or van der Waals forces calculated from
the Hamaker theory based on pairwise dipole interactions and, at a more
sophisticated level, via the continuum Lifshitz theory
\cite{lif}. The resulting DLVO \cite{is,rusasc}
theory has been very successful in describing the physics of electric double 
layers. However, the splitting of the overall interaction into a Poisson
Boltzmann static interaction and the van der Waals forces as two distinct and
independent interactions is artificial from a global view point. 
The fluctuations of the full theory of the electrostatic interactions 
generate the zero frequency van der Waals interaction. However since the 
ion distributions do not respond to high frequency fields, the non zero
frequency contributions to the van der Waals forces can be taken to be
effectively independent of the ionic distribution. The zero frequency
contributions between neutral surfaces or surfaces with absorbed 
mobile ions or dipoles can be treated and one finds that the zero
frequency van der Waals forces become screened by the presence of  
an electrolyte \cite{mani}. 
In the context of the Debye-H\"{u}ckel approximation these
van der Waals forces have a formally identical origin to the Casimir
effect, where the suppression of fluctuations of the electromagnetic field 
due to the presence of two surfaces leads to a net attractive force. The
appearance of such generalized Casimir forces in soft condensed matter
systems and many other contexts is now well established 
\cite{podo,kago,motr}. 

The theory proposed for electrolytic soap films in \cite{dese}
can in fact be solved exactly in one dimension \cite{dehose},
via path integral techniques originally developed in \cite{edle}. 
In this exact 
theory one finds that the disjoining pressure is repulsive at 
large distances but attractive contributions come into play at smaller 
inter-surface separations and can lead to a disjoining pressure isotherm
predicting a collapse to a thinner film state at a certain thickness, as
is seen in  experiments. Attractive forces in the two dimensional 
form of this model have also been found \cite{teme} at a particular critical
temperature where the model is exactly soluble.
In this paper we revisit the theory of 
\cite{dese} but in three dimensions. We use a field
theory representation of the system 
in the weak coupling limit,
equivalent to the region where the Debye-H\"{u}ckel approximation is valid, 
where the field theory is free and one may decouple the Fourier components 
of the field and apply standard path integral results. 
This allows one to incorporate a thermodynamic or energetic 
surface-charging mechanism in a straightforward way and also allows one to take into account spatially varying dielectric constants. In principle
the solution of the free field theory can be written down in terms of
a functional determinant which may be evaluated via functional techniques 
\cite{van,mani,atmini1,atmini2,atkjmi}; these functional techniques
could also be applied here, however the path integral method gives a
very compact and rapid solution to the problem. 

Recent experimental studies of the CBF-NBF transition have been carried out
on nonionic soap films made with the surfactant $C_{12} E_6$
\cite{caetal}. These
films can be stabilized in the presence of small concentrations 
of electrolyte and are thus in a region where the linearized 
Poisson-Boltzmann equation can be used for the mean field treatment and 
also where the free or Gaussian field theory we examine should be valid.

\section{The Model and Surface-Charging Mechanisms}
Here we explain the derivation of the type of model proposed in
\cite{dese,dehose} but with a generic surface-charging mechanism.
We consider the case of a monovalent electrolyte for simplicity.
In the presence of the surface, a surface charge may be generated
by two basic mechanisms. Firstly, as was taken to be the case in
\cite{dese,dehose}, there may be an affinity for one of the 
species to be on the surface. In the case of SDS, or other ionic films, this
species is the soap tail as, due to the hydrophobic effect,
the hydrocarbon tail lowers the free energy of the system by leaving the 
aqueous core and entering the air environment at the exterior of the
film. The head of the surfactant, $SO_4^-$ for SDS, is however negatively 
charged, thus leading to a surface charge. In the case of nonionic soap 
films the surface-charging mechanism is due to a difference in mobility
of the ions. Hydration effects can make one species effectively larger 
than another thus leading to a steric repulsion at the surface and hence an 
effective charge is induced by the unneutralized presence of the other species.
The enhancement of the  repulsion between the surfaces of ionic Aerosol-OT 
films due  to more  hydrated counter-ions, e.g. LiCl instead of
CsCl, has  been experimentally demonstrated in \cite{sebe}. 
Let us consider a model with one salt species (water is also partially
dissociated and indeed plays an important role in the absence of salt), the
generalization to several species is straightforward.

Let us denote by $V_1(x)$ and $V_{-1}(x)$ the effective potential for 
the cations and anions respectively due to the surfaces of the film. 
The  Hamiltonian for the electrolyte system is thus

\begin{equation}
H = {1\over 2}e\sum_{i} \psi({\bf x}_i)q_i
+ \sum_{i} V_{q_i}({\bf x}_i) 
\end{equation}
where the first term is the electrostatic energy with potential 
$\psi$, $q_i = \pm 1$  if particle $i$ is a cation/anion respectively 
and ${\bf x}_i$ its position. The second term is the 
boundary interaction. We denote by $L$ the perpendicular distance between
the two surfaces (in the direction $z$) and by $A$ the surface area of the 
film (in the plane $(x,y)$). We denote by $A\times L$ the region inside
the film and by $A\times T$ the region outside the film. The total length
of the system in the $z$ direction is denoted by $U$ and therefore
$T = U-L$. To start with, we ignore the fact that just outside
the surface one has a density of hydrocarbon tails (this will be taken into
account very simply later on). The 
calculation for a triple layer in the absence of surface 
charging and electrolyte can be found in \cite{nipa2,dorivr}.
In this two layer picture therefore, the dielectric constant
has the form $\epsilon({\bf x}) = \epsilon$ for ${\bf x}\in A\times L$
where $\epsilon$ is the dielectric constant of the electrolyte solution, which
in the dilute limit we shall take to be the dielectric concentration of water. 
Outside the film we have 
$\epsilon({\bf x}) = \epsilon_0$ (${\bf x}\in A\times T$), where
$\epsilon_0$ is the dielectric constant of air. 
The difference between the internal and external dielectric constants
means that one needs to take into account image charges and the 
zero frequency terms of the Lifshitz theory, however this
is automatically incorporated in the field theoretic 
formulation of the theory.

The electrostatic potential $\psi$ satisfies the Poisson equation

\begin{equation}
\nabla\cdot \epsilon({\bf x}) \nabla \psi({\bf x}) = -e\sum_i q_i \delta({\bf x}-{\bf x}_i)
\end{equation}
 
We follow the standard method for converting a Coulomb system to a Sine-Gordon
like field theory by performing a Hubbard-Stratonovich transformation
with an auxiliary field $\phi$ to obtain the grand canonical partition 
function

\begin{equation}
\Xi = \int d[\phi] \exp\left(S[\phi]\right)
\end{equation}

where

\begin{eqnarray}
S[\phi] = -{1\over 2} \int_{(T+L)\times A} \beta \epsilon({\bf x}) (\nabla \phi)^2
d{\bf x} 
&+& \mu \int_{L\times A} \exp\left( i \beta e \phi - \beta V_{1}({\bf x})\right)d{\bf x}
\nonumber \\
\mu \int_{L\times A} \exp\left( -ie\beta \phi - \beta V_{-1}({\bf x})\right)d{\bf x}
\label{act1}
\end{eqnarray}
and where $\mu$ is the fugacity of the cations and anions. 
It should be noticed 
that the Hubbard Stratonovich-transformation is over all space and hence
the first, kinetic, term of the action is an integral over all space
denoted $ (T+L)\times A$ and the second two terms are the interaction terms
restricted to the film region $L\times A$. We notice that the 
functional integral gives the ion/ion interaction upon 
performing the integral, but in addition there is a term
\begin{equation}
\int d[\phi] \exp\left({\beta \over 2}\int \phi \nabla\cdot \epsilon({\bf x}) \nabla\phi d{\bf x}\right)
=\left(\det(- \nabla\cdot \epsilon({\bf x}) \nabla )\right)^{-{1\over 2}}
\label{deter}
\end{equation}
This term is however the zero frequency contribution coming from the 
Lifshitz theory and in fact should be there. This term naturally arises when  
one considers the full Quantum Electrodynamics (QED) of the system. If 
one fixes the positions of the ions and ignores the magnetic 
part of the Lagrangian, the time independent (zero
Matsubara frequency) purely electrostatic part of the Lagrangian
is \cite{fehi}
\begin{equation}
{\cal L}[\psi] = {1\over 2}\int \epsilon({\bf x}) (\nabla \psi)^2 d{\bf x} - e \sum_{i} q_i 
\psi({\bf x}_i)  \label{action}
\end{equation}
hence the thermal field theory for the field $\psi$ has a partition function
\cite{balo}
\begin{equation}
Z = \int d[\psi] \exp\left(\beta {\cal L}[\psi]\right)
\label{eqzqeds}
\end{equation}
If one now takes a classical trace over the ion positions  and 
passes to the grand canonical ensemble, one finds the expression for the
grand potential above after changing the axis of the functional integration
via $\psi \to -i \phi$. If one wishes to take into account the non zero 
frequency Lifshitz terms one proceeds as above but keeping the full
QED action and introducing the bosonic Matsubura frequencies $\omega_n
= 2\pi n/\beta \hbar$ \cite{balo}.
Hence the grand partition function $\Xi$ contains the 
ionic interactions and zero frequency van der Waals 
contributions in the system.  
We define a Stern layer to be the region where either or both of  
the potentials $V_{\pm 1}$ are non zero and take the width of this region to
be $\delta$. The action then has the form
 
\begin{eqnarray}
S[\phi] = &-&{1\over 2} \int_{(T+L)\times A} \beta \epsilon({\bf x}) (\nabla \phi)^2
d{\bf x} 
+ 2\mu \int_{L\times A} \cos\left( \beta e \phi\right)d{\bf x}
\nonumber \\
&+&\mu \int_{[0,\delta]\times A} \exp\left( i\beta e \phi\right) 
\left(\exp\left(- \beta V_{1}({\bf x})\right) - 1\right)d{\bf x}
+ \mu \int_{[L-\delta,L]\times A} \exp\left( i\beta e \phi\right) 
\left(\exp\left(- \beta V_{1}({\bf x})\right) - 1\right)d{\bf x}
\nonumber \\
&+&\mu \int_{[0,\delta]\times A} \exp\left( -i\beta e \phi\right) 
\left(\exp\left(- \beta V_{-1}({\bf x})\right) - 1\right)d{\bf x}
+ \mu \int_{[L-\delta, L]\times A} \exp\left( -i\beta e \phi\right) 
\left(\exp\left(- \beta V_{-1}({\bf x})\right) - 1\right)d{\bf x}
\label{act2}
\end{eqnarray}
We now take $\delta$ to be small and take the limit $\delta \to 0$ choosing
\begin{equation}
\left(\exp\left(- \beta V_{ \pm 1}({\bf x})\right) - 1\right)
  \to \mu^*_\pm (\delta(z) + \delta(L-z)) \label{limit}
\end{equation}
where $\delta(z)$ is the one dimensional Dirac delta function.
Clearly $\mu^*_{\pm}$ is positive/negative for $V_{\pm 1}$ negative 
(attraction 
of the species towards the surface)/ positive (repulsion of the species from
the surface). Note that dimensionally $[\mu^*_{\pm}] = [\mu]\ [\delta]$.
In this simplified limit, the action is now

\begin{eqnarray}
S[\phi] = -{1\over 2} \int_{(T+L)\times A} \beta \epsilon({\bf x}) (\nabla \phi)^2
d{\bf x} 
&+& 2\mu \int_{L\times A} \cos\left( \beta e \phi\right)d{\bf x}
\nonumber \\
\ &+& \mu_+^*  \int_{L\times A} (\delta(z) + \delta(L-z))\exp\left(i\beta e \phi\right)
d{\bf x} \nonumber \\
&+& \mu_-^*  \int_{L\times A} 
(\delta(z) + \delta(L-z))\exp\left(-i\beta e \phi\right)\ d{\bf x}
\label{actf}
\end{eqnarray}
In the weak coupling limit we proceed by expanding the action to quadratic 
order in all terms, yielding the Gaussian action
\begin{eqnarray}
S_G&&[\phi] ~=~ -{1\over 2} \int_{(T+L)\times A} \beta \epsilon(z) (\nabla \phi)^2
d{\bf x} - {1\over 2}\int_{L\times A} \beta \epsilon m^2 \phi^2 \ d{\bf x}
\nonumber \\ +~
 i\lambda &&\left( \int_{z=0} \phi \ dxdy + \int_{z=L} \phi \ dxdy
\right) - {1\over 2}\beta\epsilon m^2 \gamma \left( \int_{z=0} \phi^2 \ dxdy + \int_{z=L} \phi^2 \ dxdy
\right) + 2\mu A L + 2(\mu^*_+ +\mu^*_-)A
\label{actl} 
\end{eqnarray}
Here $m = \sqrt{2\mu e^2 \beta/\epsilon}$ corresponds to the Debye mass (the 
inverse Debye length) in the dilute, weak coupling limit, as in this limit 
$\mu \approx \rho$, where $\rho$ is the bulk electrolyte concentration.
In addition we have 
\begin{equation}
\lambda = e\beta(\mu^*_+ - \mu^*_-)~\equiv~m^2 \bar{\g}\;,~~~\barg =  {\mu^*_+ - \mu^*_-\over 2 \mu}
~~~\gamma = {\mu^*_+ + \mu^*_-\over 2 \mu} \label{gammas}
\end{equation}
and we recall $\epsilon(z) = \epsilon$ for $z\in [0,L]$ and is equal to
$\epsilon_0$ elsewhere. 
We now express $\phi$ in terms of its Fourier decomposition in the 
${\bf r} = (x,y)$ plane.

\begin{equation}
\phi({\bf r},z) = {1\over \sqrt{A}}\sum_{\bf p} {\tilde \phi}({\bf p}, z)
\end{equation}
where if $A = l\times l$ imposing periodic boundary conditions yields
${\bf p} \in {2\pi \over l}(n_x,n_y)$ with $(n_x,n_y) \in {Z}^2$.
In terms of the fields ${\tilde \phi}$, the action decouples and one obtains
\begin{equation}
S_{G} = S_{0}[{\tilde \phi}(0)] + \sum_{{\bf p}\neq 0} S_{{\bf p}}[{\tilde \phi}({\bf p})] + 2\mu AL + 2(\mu^*_+ +\mu^*_-)A
\end{equation}
where
\begin{eqnarray}
S_{0}[{\tilde \phi}(0)] &=& -{1\over 2} \int_{T+L} 
\beta \epsilon(z) \left({\partial {\tilde \phi}(0,z)
\over \partial z}\right)^2   \ dz- {1\over 2} \beta \epsilon m^2
\int_{L} {\tilde \phi}^2(0,z) \ dz \nonumber \\
&+& i \lambda \sqrt{A} \left( {\tilde \phi}(0,0) + {\tilde \phi}(0,L)\right)
- {1\over 2}\beta \epsilon \gamma m^2  
\left( {\tilde \phi}^2(0,0) + {\tilde \phi}^2(0,L)\right)
\end{eqnarray}
and 
\begin{eqnarray}
S_{{\bf p}}[{\tilde \phi}({\bf p})]&=& -{1\over 2} \int_{T+L} 
\beta \epsilon(z) \left({{\partial {\tilde \phi}({\bf p},z)
\over \partial z}}{{\partial {\tilde \phi}({\bf -p},z)
\over \partial z}} + {\bf p}^2 {\tilde \phi}({\bf p},z){\tilde \phi}({\bf -p}
,z)
 \right) dz
\nonumber \\ &-&
{1\over 2} \beta \epsilon m^2\gamma
\int_{L} {\tilde \phi}({\bf p},z){\tilde \phi}({\bf -p},z) \ dz
\end{eqnarray}
The system therefore decomposes into a system of simple harmonic 
oscillators and 
\begin{equation}
\Xi = {\cal N}\exp\left( 2 \mu A L + 2 A (\mu_+^* +\mu_-^*)\right) 
\int d[\tilde{\phi}(0)] \exp[S_0] \prod_{{\bf p}\neq 0}
d[\tilde{\phi}({\bf p})] \exp[S_{\bf p}]
\end{equation}
where in the following ${\cal N}$ will be used to note  $L$ independent 
normalization factors.
Each oscillator (labeled by {\bf p}) has a time dependent Hamiltonian
\begin{equation}
H_{\bf p} = -{1\over 2 M(z)}{d^2 \over dX^2} + 
{1\over 2}M(z) \omega^2({\bf p}, z)X^2 \label{hamiltonian}
\end{equation}
where $M(z) = \beta \epsilon(z)$ and $\omega({\bf p},z) =
\sqrt{{\bf p^2} + m^2}$ for $z\in [0,L]$ and $\omega({\bf p},z) =
|{\bf p}| = p$ for $z\notin [0,L]$. Here $X$ represents the field $\phi({\bf p})$ 
and $z$ corresponds to the temporal coordinate. The mode ${\bf p} = 0$ is
slightly more complicated due to the presence of a linear source term. 
One may write
{\bacol
\bea
\int d[\tilde{\phi}(0)]&& \exp[S_0]~= \nn\\
&&{\cal N}_0 {\rm Tr} \ \exp\left( -(U-L)H_E \right)
\exp\left(i\lambda \sqrt{A}X - {1\over 2} \beta \epsilon m^2 \gamma X^2\right)
\exp\left( -L H_F \right) \exp\left(i\lambda \sqrt{A}X - {1\over 2} \beta \epsilon m^2 \gamma X^2 \right)
\label{mode0}
\eea
}
where $H_E$ is the Hamiltonian outside the film with mass 
$M_E = \beta \epsilon_0$ and $\omega_E = 0$ and $H_F$ is the Hamiltonian
in the film with $M_F = \beta \epsilon$ and $\omega_F = m$
(subscripts $E$ and $F$ will be used to denote the masses and frequencies
external and internal to the film respectively). In the limit where
$U \to \infty$ only the eigenstate of $H_E$ of lowest energy survives the
thermodynamic limit (we shall see later that this is compatible with
the constraint of electro-neutrality within the film at mean field level).
We recall that the ground-state wave function of the simple harmonic
oscillator is given up to a normalization by
\begin{equation}
\psi_{0}(X,M, \omega) = \exp(-{M\omega X^2\over 2})
\end{equation}
and the corresponding ground-state energy is ${1\over 2}\omega$. In the 
case where ${\bf p} = 0$ one has the ground-state wave function $\psi_{0}(X,M_E,0) = 1$. 
Using this in Eq. (\ref{mode0}) thus yields 
{\bacol
\begin{eqnarray}
\int d[\tilde{\phi}(0)]&& \exp[S_0]~= \nonumber \\ 
&& {\cal N}_0 
\int \exp\left(i\lambda \sqrt{A} X - {1\over 2} M_F  m^2 \gamma X^2\right)
K(X,Y,L,\omega_F,M_F)
\exp\left(i\lambda \sqrt{A} Y - {1\over 2} M m^2 \gamma Y^2
\right) \ dX dY \label{P=0}
\end{eqnarray}
}
where $K(X,Y,L,\omega,M)$ is the Feynman kernel for the simple Harmonic 
oscillator \cite{fehi} and is given by
\begin{equation}
K(X,Y,L,\omega,M) = \left( {M \omega \over 2 \pi \sinh(\omega L)}\right)^{1\over 2} \exp\left(-{1\over 2} M \omega \coth(\omega L)(X^2 + Y^2 - 2XY {\rm sech}(\omega L))\right)
\end{equation}
Performing the above integral yields a term extensive in $A$ coming from the
linear source term and a non extensive fluctuation term, which we shall absorb
into the normalization term, yielding
\begin{equation}
{\ln\left(\int d[\tilde{\phi}(0)] \exp[S_0]\right)\over A} =
-{\lambda^2 ( m\gamma\sinh(mL) + \cosh(mL)  + 1) \over 
\beta \epsilon m ((m^2 \gamma^2 + 1)\sinh(mL) 
+ 2 m\gamma \cosh(mL))}
\end{equation}
This term can be simplified slightly giving the zero momentum contribution
to the grand potential per area of film. Writing $J = -\ln(\Xi)/(\beta A)
= J_c + \sum_{\bf p} J_{{\bf p}}$, where
\begin{equation}
J_c = -{2\over \beta}\left(\mu L + \mu_+^* + \mu_+^*\right)
\end{equation}
is ideal part of the grand potential, we find
\begin{equation}
\beta J_{0} = 
2m\mu\barg^2{\cosh\left( {mL\over 2}\right)
\over \sinh\left( {mL\over 2}\right) + m\gamma\cosh\left( {mL\over 2}\right)}
\label{S0}\end{equation}
In fact we shall see later, as should be expected,
that this term is the mean field contribution to the theory.

The non zero momentum modes only contribute to the fluctuations but summed 
together give an extensive contribution. The contribution to the
grand partition function of one of these modes is
{\bacol
\begin{eqnarray}
\int d[\tilde{\phi}({\bf p})]&&\exp[S_{\bf p}] =\nonumber \\
&&{\cal N}_{p} {\rm Tr}\;\left[\;\exp(-{M_E\omega_E X^2\over 2}) \exp\left( -(U-L)H_E \right)
\exp\left(- {1\over 2} \beta \epsilon m^2 \gamma X^2\right)\right. \nn\\
&&~~~~~~~~~~\left.\exp\left( -L H_F \right) \exp\left(- {1\over 2} \beta \epsilon m^2 \gamma X^2 \right)
\exp(-{M_E\omega_E X^2\over 2})\right]
\label{modep}
\end{eqnarray}
}
Note that there is no under-counting in this contribution as the field 
$\phi$ was real and hence $\overline{\tilde{\phi}}(p) =  {\tilde{\phi}}(-p)$.
One thus obtains
\begin{eqnarray}
&\ &\int d[\tilde{\phi}({\bf p})] \exp[S_{\bf p}] = {\cal N}_{{\bf p}}\exp
\left(-{U-L\over 2}\omega_E\right) \times
\nonumber \\ 
&\ & 
\int \exp\left( -{M_E\omega_E X^2\over 2} - {1\over 2} M_F  m^2 \gamma X^2\right)
K(X,Y,L,\omega_F,M_F)
\exp\left(-{M_E\omega_E Y^2\over 2}- {1\over 2} M_F m^2 \gamma Y^2
\right) \ dX dY
\end{eqnarray} 

This yields an $L$ dependent contribution to the grand potential per unit 
area of
\begin{equation}
J_{{\bf p}} = -{1\over A\beta}\left[ {1\over 2}L( p - 
\sqrt{p^2 + m^2})
-{1\over 2} \ln\left(1 - \left( {\epsilon_0 p + \epsilon m^2 \gamma 
-\epsilon\sqrt{p^2 + m^2}
\over   \epsilon_0 p + \epsilon m^2 \gamma +\epsilon\sqrt{p^2 + m^2}}\right)^2 \exp(-2L \sqrt{p^2 + m^2})\right)\right]
\label{jf}
\end{equation}
The total contribution from the modes ${\bf p} \neq 0$ is the grand potential
associated with the static electrostatic fluctuations, or the zero
frequency van der Waals force, hence we write $J_{vdW} = \sum_{{\bf p} \neq 0}
J_{{\bf p}}$. We should remark however, given the 
unifield treatment used here, this notation is slightly arbitary as
with the definition used here $J_{vdW}$ contains 
terms coming from the ionic fluctuations.  
The first term in the expression (\ref{jf}) 
will clearly lead to an ultraviolet 
divergence in the grand potential, however the disjoining 
pressure $P_d(L)$ is given by the difference between the 
film and bulk pressures
\ben
P_d(L) ~=~P(L) - P_{{\rm bulk}} ~=~ -{\partial J(L)\over \partial L} + \lim_{L\to \infty} J(L)/L 
       ~=~ -{\partial J^*(L)\over \partial L} \label{pdis}
\een
After summing over all values of ${\bf p}$, by passing to the continuum 
for large $A$, we obtain the divergence free result 
\begin{eqnarray}
\beta J^*(L) &=& {1\over 4 \pi}\int_0^\infty dp \ p \ln\left( 1 - 
\left( {\epsilon_0 p + \epsilon m^2 \gamma  -\epsilon\sqrt{p^2 + m^2}
\over   \epsilon_0 p +\epsilon m^2 \gamma + \epsilon\sqrt{p^2 + m^2}}\right)^2 
\exp(-2L \sqrt{p^2 + m^2})\right) \nonumber \\
&+& 2m\mu\barg^2{\cosh\left( {mL\over 2}\right)
\over \sinh\left( {mL\over 2}\right) +m\gamma\cosh\left( {mL\over 2}\right)}
\nonumber \\ &=& \beta J^*_{vdW}(L) + \beta J_0(L)
\label{jstar}
\end{eqnarray}
From Eq. (\ref{pdis}) it is clear that $J^*(L)$ is the effective interaction 
potential between the two surfaces.

The bulk grand potential contains the extensive contribution which itself contains the
divergence. By taking the difference we have concentrated on the part relevant to the
calculation of the disjoining pressure. However, it is instructive to remark that the
extensive term in Eq. (\ref{jf}) is the term in the bracket that is explicitly proportional to $L$. 
It can be shown that on integration this term does indeed give the standard Debye-H\"{u}ckel
expression for the pressure. It is, in fact, the one-loop contribution similar to that
explicitly discussed in the 1D model in \cite{dehose}. The difference in the 3D case is that
the integral is divergent and so must be renormalized. The divergence for this integral
is due to the self-energy of the charge distribution included in the original
definition of the partition function. It can be removed by imposing the renormalization
condition the the fugacity $\mu$ be chosen so that the density is given by
\be
\rho~=~-{\mu \over \beta}{\partial\log{\Xi} \over \partial\mu}~.
\ee
$\mu$ is then a bare quantity which is a function of the Ultra-Violet cutoff. It can be
shown that when the appropriate one-loop calculations are performed the bulk pressure
is finite and is given by
\be
\beta P_{\rm bulk}~=~\rho - {m^3 \over 24\pi}~.
\ee
Where they are important the UV divergences will ultimately be regulated by the finite size
of the ions. However, these contributions will be corrections and it will be possible
to estimate their importance using perturbative methods. The details will be postponed to a 
further paper which will report on the perturbative approach in general. The renormalization of 
$\mu$ is not of relevance to what follows but is central to higher-order corrections.

\section{Various Limits}
To demonstrate the generality of the formalism here and its compatibility
with well established results in the literature derived by other means,
we shall consider various limits of the formulas derived above. 
\subsection{Mean Field Limit}
The mean field theory for the action (\ref{actl}) is obtained 
from the equation
\begin{equation}
{\delta \over \delta \phi({\bf x})}S_G[\phi] = 0
\end{equation}
This yields the equation
\begin{equation}
\beta\nabla\cdot \epsilon \nabla \phi - m^2 \beta \epsilon \phi
+ i\lambda(\delta(z) + \delta(z-L)) - \beta\epsilon m^2\gamma\phi
(\delta(z) + \delta(z-L)) = 0
\end{equation}
within the film. Outside the film one has
\begin{equation} 
\beta\nabla\cdot \epsilon_0 \nabla \phi = 0 \label{mfout}
\end{equation}
As is usual for Euclidean Sine-Gordon field theories, the physical 
saddle point is imaginary, one writes $\phi = i\psi$ and it turns
out that $\psi$ is the mean field electrostatic potential, this is also
clear from the static QED formulation (\ref{eqzqeds}). In addition the
mean field solution depends only on the coordinate $z$. 
From Eq. (\ref{mfout}) one finds that outside the film $d\psi/dz = 0$ which
is simply the condition of electro-neutrality in the film. Inside the film
(strictly away from $z=0$ and $z=L$) one has therefore
\begin{equation}
\beta \epsilon \left({d^2\over dz^2}\psi - m^2 \psi\right) = 0
\end{equation}
which is simply the linearized Poisson Boltzmann equation.
The solution to this equation which gives a potential symmetric about the
films midpoint $L/2$ is
\begin{equation}
\psi = C \cosh\left(m(x - {L\over 2})\right)
\end{equation}
Integrating the mean field equation between $z= 0^-$ and $z= 0^+$, and
using the condition of electro-neutrality, one finds
\begin{equation}
-\beta \epsilon {d \psi\over dz}|_{z= 0^+} - \beta \epsilon m^2\gamma\psi(0)
+ \lambda = 0
\end{equation}
This allows one to solve for $C$ giving
\begin{equation}
C = {\lambda \over \beta\epsilon m\left(\sinh({mL\over 2}) + m\gamma
\cosh({mL\over 2})\right)}
\end{equation}
Substituting this mean field solution into the expression (\ref{actl}) for 
$S_G$, yields the mean field action which is exactly the expression
(\ref{S0}) obtained from the zero momentum contribution in the previous
analysis.
\subsection{No electrolyte no surface-charging}
This amounts to calculating the zero van der Waals forces across a
slab of dielectric constant $\epsilon$ while the dielectric constant
of the external media  is given by $\epsilon_0$. Here one has

\begin{equation}
J^*(L) = {k_B T\over 4 \pi}\int_0^\infty dp \ p \ln\left( 1 - 
\left( {\epsilon_0  -\epsilon
\over   \epsilon_0 + \epsilon}\right)^2 
\exp(-2Lp)\right)
\end{equation}
The integral above is easily evaluated by expanding the logarithm leading to
\begin{equation}
J^*(L) = -{k_B T\over 16 \pi L^2} \sum_{n=1}^{\infty} {1\over n^3} 
\left( {\epsilon - \epsilon_0 \over \epsilon + \epsilon_0}\right)^{2n}
\end{equation}
This yields the attractive, Casimir like, disjoining pressure 
\cite{mani,atkjmi}
\begin{equation}
P_d(L) = -{k_B T\over 8 \pi L^3} \sum_{n=1}^{\infty} {1\over n^3}\left( {\epsilon - \epsilon_0 \over \epsilon + \epsilon_0}\right)^{2n}
\end{equation}

\subsection{No surface-charging}
If one considers the limit where there is no interaction between the ions
and the surface one has $\mu_+^* = \mu_-^* = 0$ and hence $\gamma = 0$. This 
is the case of an electrolyte confined by two neutral surfaces (in both 
the electrostatic and chemical sense) but with a medium of dielectric 
constant $\epsilon_0$ out side the film and dielectric 
constant $\epsilon$ inside the film. One thus obtains 
\begin{equation}
J^*(L) = {k_B T\over 4 \pi}\int_0^\infty dp \ p \ln\left( 1 - 
\left( {\epsilon_0 p -\epsilon\sqrt{p^2 + m^2}
\over   \epsilon_0 p + \epsilon\sqrt{p^2 + m^2}}\right)^2 
\exp(-2L \sqrt{p^2 + m^2})\right)
\end{equation}
This thus recovers the result of \cite{mani} demonstrating the 
screening of the zero frequency van der Waals interaction in the 
presence of an electrolyte.

\subsection{Localized Surface Charge Fluctuations}
It is easy to see that the term proportional to $\gamma$ in the action
(\ref{actl}) is a term proportional to the surface charge fluctuation 
about its mean value. The effect of surface charge fluctuations in the limit
where the counterions are highly localized about the surface was considered 
in \cite{atkjmi} and \cite{pisa}. In \cite{atkjmi} the problem was considered
in the strong and weak coupling Debye-H\"{u}ckel regime, in this regime
the electrostatic fluctuations lead to the computation of a mathematically 
identical determinant in both  \cite{atkjmi} and \cite{pisa} and the 
electrostatic fluctuations in our model also have the same mathematical 
structure. The differences between \cite{atkjmi,pisa} and our model are in
terms of the physical interpretation at this level.
In the case of non varying  dielectric constant (for instance the case of 
two membranes interacting in water) the contribution to the 
disjoining pressure coming from the electrostatic fluctuations 
or the zero frequency van der Waals term $P_d^{(vdW)}(L)$
is given by  taking $m=0$ within the film. 
Immediately one obtains

\begin{equation}
P_d^{(vdW)}(L) = -{k_B T \over 2\pi} \int dp { p^2 \over \left[
\exp(2p L) (1 + p {\overline\lambda})^2 -1\right]}\label{eqfluc}
\end{equation}
where $ {\overline\lambda} = {2/ m^2\gamma}$. The expression 
(\ref{eqfluc}) is up to the definition of ${\overline\lambda}$ the
same as obtained in and \cite{atkjmi} and \cite{pisa}.

In \cite{atkjmi} the problem of localized surface-charging but with varying 
dielectric constant, that outside the film being different to that within 
the film, was also considered. In \cite{mepist} the model of 
\cite{pisa} was studied incorporating the same  dielectric 
constant variation. However at a mathematical level the calculation of the 
electrostatic fluctuations in \cite{atkjmi},\cite{mepist} and here 
is again identical.
Here we find that the contribution from the electrostatic fluctuations is
\begin{equation}
P_d^{(vdW)}(L) = -{k_B T \over 2\pi} \int dp { p^2 \over \left[
\exp(2p L) \left( {(\epsilon + \epsilon_0)p + \epsilon m^2 \gamma
\over(\epsilon - \epsilon_0)p + \epsilon m^2 \gamma}\right)^2  -1\right]}
\label{eqflucvd}
\end{equation}
This equation agrees with the result of \cite{atkjmi} (in the Debye
H\"uckle regime) which is obtained there by
calculating the contribution of the ionic fluctuations and then adding the 
contribution coming from the zero frequency Lifshitz terms. In 
\cite{mepist} only the ionic fluctuations were included and hence
the result differs from ours and that of \cite{atkjmi} by the zero frequency 
Lifshitz term. 

\subsection{Including the Non-zero Frequency van der Waals Contribution}
Let us mention that if one wanted to now take into account 
non-zero frequency contributions from the Lifshitz theory, which are 
in principle uncoupled or very weakly coupled to the static ionic fluctuations 
one should use a total potential 

\begin{equation}
J_T(L) = J^*(L) + \sum_{\omega\neq 0}J_{{\rm Lif}}(w,L)
\end{equation}
where $J_{\rm{Lif}}(\omega,L)$ is the contribution to the dispersion potential
coming from the Matsubara frequency $\omega$ in the absence of electrolyte. 
As mentioned previously, in the formulation here, the zero frequency 
Lifshitz and ionic components are treated together in the 
grand partition function $\Xi$.  

\section{The Triple Layer}
As mentioned earlier, the change in the dielectric constant due the 
the presence of the hydrocarbon layer formed just outside the 
surface of the soap film by the surfactant should also be taken 
into account. If one considers this to be a region of size $h$ 
at either interface, but inaccessible to the aqueous solution one
now has an different Hamiltonian $H_H$ (where the subscript $H$ denotes
the hydrocarbon region) in these regions characterized,
for a mode ${\bf p}$, by $\omega_H = p$ and $M_H = \beta \epsilon_1$,
where $\epsilon_1$ is the dielectric constant of this hydrocarbon
region (strictly this is a mixture of air/hydrocarbon chains).
Using the formalism developed here one finds that the mode ${\bf p}$ 
yields a contribution 
\begin{eqnarray}
&\ &\int d[\tilde{\phi}({\bf p})] \exp[S_{\bf p}] = {\cal N}_{{\bf p}}\exp
\left(-({U-L-2h\over 2})\omega_E\right) \times
\nonumber \\ 
&\ &
\int \exp\left( -{M_E\omega_E X^2\over 2}\right)K(X,Y,h,\omega_H,M_H)
 \exp\left(- {1\over 2} M_F  m^2 \gamma Y^2\right) \times \nonumber \\
&\ &K(Y,Z,L,\omega_F,M_F)
\exp\left(- {1\over 2} M_F m^2 \gamma Z^2\right)K(Z,W,h,\omega_H,M_H)
\exp\left(-{M_E\omega_E W^2\over 2}\right) dX dY dZ dW
\end{eqnarray}
to the grand partition function.

The ${\bf p}=0$ mode or mean field contribution is unchanged by the 
addition of the hydrocarbon layer and after some 
algebra on finds that

\begin{eqnarray}
\beta J^*(L) &=& {1\over 4 \pi}\int_0^\infty dp \ p \ln\left( 1 - 
\left({  \epsilon_1 B(p)p
+\epsilon (m^2\gamma -\sqrt{p^2 + m^2}) 
\over    \epsilon_1 B(p) p 
+\epsilon (m^2\gamma +\sqrt{p^2 + m^2})
}\right)^2
\exp(-2L \sqrt{p^2 + m^2})\right) \nonumber \\
&+& 2m\mu\barg^2 {\cosh\left( {mL\over 2}\right)
\over\sinh\left( {mL\over 2}\right) + m\gamma\cosh\left( {mL\over 2}\right)}
\label{jft}
\end{eqnarray}
where
\begin{equation}
B(p) = {1 + \Delta \exp(-2ph)\over 1 - \Delta \exp(-2ph)}
\end{equation}
with
\begin{equation}
\Delta = {\epsilon_0 - \epsilon_1\over \epsilon_0 + \epsilon_1}
\end{equation}
Note that if $\epsilon_0 = \epsilon_1$ we recover the double layer
result obtained earlier. In the limit $h\to 0$ we obtain 
$\epsilon_1 B(p) \to \epsilon_0$ and when $h\to \infty$ one has
$\epsilon_1 B(p) \to \epsilon_1$ as it should.  A key, physically illuminating
step, in the algebra mentioned above is the evaluation of the term
\begin{equation}
\psi^*(Y) = \int \exp(-{1\over 2} M_E \omega_E X^2) K(X,Y,h,M_H,\omega_H) dX
\end{equation}
One finds that
\begin{eqnarray}
\psi^*(Y) &=& \left( M_H \omega_H\over M_E \omega_E \sinh(\omega_H h)
 +  M_H \omega_H \cosh(\omega_H h)\right)^{1\over2} \nonumber
\\
&\ &\exp\left(-{1\over 2}Y^2\left({ M_E \omega_E M_H \omega_H \coth(\omega_H h)
+ M_H^2 \omega_H^2 \over M_E \omega_E + M_H \omega_H \coth(\omega_H h)}
\right)\right)
\end{eqnarray}
The normalization here is unimportant for the calculation of the disjoining 
pressure (as $h$ is taken to be fixed). 
We can now interpret $\psi^*$ as the new effective
ground-state wave function associated to the mode ${\bf p}$ entering the 
film. As there is no electrolyte in the hydrocarbon region one has
that $\omega_H = \omega_E = p$ (the case where electrolyte was 
present could be easily handled by the same formalism) and hence
one finds that the effective or renormalized ground-state wave function
entering the film is
\begin{equation}
\psi_0(X) = \exp\left(-{1\over 2}X^2 \omega_H M_H
\left({ M_E \coth(p h)
+ M_H \over M_E + M_H \coth(p h)}\right)\right)
\end{equation}
Clearly this leads to a renormalization of $M_H$ with respect to 
what it would be if the region of hydrocarbons was in fact of infinite size,
and consequently one finds a $p$ dependent renormalization
of the dielectric constant $\epsilon_1$ as
\begin{equation}
\epsilon^R_1(p) = \epsilon_1 B(p) = \epsilon_1\left({ \epsilon_0 \coth(p h)
+ \epsilon_1 \over \epsilon_0 + \epsilon_1 \coth(p h)}\right)
\end{equation}
which is easily seen to be in agreement with the definition of $B(p)$ above.
This calculation can easily be generalized to an arbitrary number 
of electrolyte free layers simply by calculating recursively the value
of $B(p)$ generated by the composition of all the layers before arriving
at the penultimate layer before the film. In the case of no surface interaction
and no electrolyte this result agrees with the calculation of the 
zero frequency contribution to the van der Waals force in
\cite{nipa1,dorivr}.

In what concerns the the momentum dependence of $\epsilon^R(p)$
one finds that

\begin{eqnarray} \epsilon^R(p) &\approx & \ \epsilon_0\ {\rm for} \  
{1\over p} \gg h \\
\epsilon^R(p) &\approx & \ \epsilon_1\ {\rm for} \ 
{1\over p} \ll h
\end{eqnarray}
Hence the long wave length modes entering the film behave as if
the hydrocarbon layer was not there and the short wave length
modes behave as if there was only the hydrocarbon layer present. 
\section{Behavior of the Disjoining Pressure}
In this section we shall examine how the disjoining pressure is 
effected by changing the various physical parameters of the theory.
We write
\begin{equation}
P_d = P_d ^{(0)} + P_d^{(vdW)}
\end{equation}
where the first term is the mean field or zero mode contribution 
$P_d^{(0)} = -\partial J_0/ \partial L$ and
$P_d^{(vdW)} = - \partial J^*_{vdW}/\partial L$ is the zero 
frequency van der Waals contribution coming from 
the modes ${\bf p}$ such that $|{\bf p}| \neq 0$.
\subsection{The Mean Field Contribution}
The mean field contribution to the disjoining pressure
is
\begin{equation}
P_d^{(0)}(L) = {\mu m^2\barg^2 \over \gb(\sinh({mL\over 2}) + m\gamma
\cosh({mL\over 2}))^2}
\end{equation}
which is clearly always positive. This repulsive component
is generated by an effective surface charge which is present when 
$\mu_+^* - \mu_-^* \neq 0$, that is to say there is an asymmetry between the
cation/anion  affinities or repulsions at the surface. At large 
inter-surface separations one finds
\begin{equation}
P^{(0)}_d(L) \approx {4\mu m^2\barg^2 \over \beta \mu (1 + m\gamma)^2}\exp(- mL)
\end{equation}
which has the standard Poisson-Boltzmann exponential decay with the  
characteristic length scale of the Debye length $l_{D} = 1/m$. If $\gamma$ is negative $\gamma = - \gamma'$ (with $\gamma' > 0$) the magnitude of the
repulsion is enhanced. Here there is a critical value $L_c$ where
the mean field component to the disjoining pressure diverges. One finds
that
\begin{equation}
L_c = {1\over m}\ln\left({1 + m\gamma'\over 1 - m\gamma'}\right)
\end{equation}
The divergence here is not physical as it can be avoided by keeping
higher order terms in the expansion of the surface terms in the full action 
(\ref{actf}). However it does indicate an enhancement of the repulsion 
due to the presence of a Stern layer of effective length
$\delta' = \gamma'$ from the considerations in section (II.).
If the theory were applicable for large $L$ and $L_c$ were small, then 
one  finds
\begin{equation}
L_c \approx 2 \delta'
\end{equation}
that is to say that the pressure should rise rapidly when the two 
effective Stern layers come into contact, thus giving a large repulsive
term in the disjoining pressure before the surfactant surfaces actually come 
into contact. This image is valid in the range where $m\delta' \ll 1$, that is
to say the width of the Stern layer is much smaller that the Debye length.
It is clear that the limit taken in Eq. (\ref{limit}) is only valid in this
case. In the case where $m\delta' \geq 1$ one must treat the Stern layer
as a continuum and introduce two new layers of finite thickness.

In the case $\gamma > 0$ the magnitude of the repulsive part of the 
disjoining pressure is decreased as $\gamma$ increases. At $L= 0$ one finds
the disjoining pressure

\begin{equation}
P_d^{(0)}(0) = \mu k_B T \left({ \barg \over \g}\right)^2
\end{equation}

Therefore at small interface separations one needs to incorporate steric
repulsion between the surfactants in the two surfaces to prevent a collapse
to a zero film thickness.  
\subsection{The Zero Frequency Van der Waals Contribution}
As mentioned previously, the presence of an electrolyte in the film
leads to a screening of the zero frequency van der Waals interaction
and simple expressions for $P_d^{(vdW)}$ as a function of $L$ do not exist
due to the presence of a second length scale the Debye length ($=1/m$). 
For simplicity we consider just the bilayer model. In the case
$L \gg 1/m$, the disjoining pressure is dominated by modes such that
$p \ll 1$ and one obtains to leading order
\ben
P_d^{(vdW)}~\sim~-{k_BTm^3 \over 4\pi mL}\left( {m \gamma -1 \over m \gamma +1}
\right)^2\;\exp(-2mL)\;.\label{hamasy}
\een
We see that the prefactor controlling the strength of the long-distance 
attraction depends crucially on the value of $\gamma$ and hence the
surface-charging mechanism. The exponential decay is however 
twice as rapid as that of the mean field contribution, meaning that
in the thick film regime the zero frequency van der Waals attraction
is dominated by the mean field repulsion term.

In the limit $L \ll 1/m$ one finds  
\ben
 P_d^{(vdW)}~\sim~-{k_BT \over L^3}\sum_{n=1}^\infty
\;{1 \over n^3}\bb {\e-\e_0 \over \e+\e_0}\eb^{2n}
 ~~~~~~L \ll 1/m~.
\label{pvdwasym}
\een
This expression is independent of $\gamma$ and 
consequently coincides with that 
given by in \cite{mani}. Hence we see that it is only in the regime 
of very thin films
that the zero frequency van der Waals force takes the Hamaker form
$P_d^{(vdW)}~\sim -1/L^3$.

\section{Orders of Magnitude and Comparison with Experimental Data}

The theory presented so far is the linearized version which is
equivalent to the Gaussian approximation and corresponds to the free
field formulation.  We should expect that a detailed comparison with
experiment will not be wholly successful since we have yet to include
non-linear effects. However, it is instructive to show qualitatively
how well the model performs and to this end we study one case for
which the parameters are typical of experiments. We consider a soap
film in air consisting of ionic liquid bounded by two thin hydrocarbon
layers of hydrophobic surfactant. We idealize the model to be a
bilayer system where we do not account for the non-zero thickness of
the hydrocarbon layers. The liquid is composed of water with a 
dissolved salt, such as NaCl. The hydrophobic surfactant is of one
charge only and so we set $\mu^*_+ \equiv \mu^*,~\mu^*_- = 0$. The
fugacity $\mu$ of the cations and anions is chosen by the experimenter
and determines the Debye mass $m$ through $m^2 = 2\mu e^2/kT$. For the
free energy we use Eq. (\ref{jf}) which we recast in terms of
dimensionless variables as 
\bea F_0(l)&=&-{\cosh(l/2)\over
\sinh(l/2)+\a\cosh(l/2)}\nn\\ F_{(vdW)}(l)&=&{1 \over 4\pi}\int dk\;k
\ln\bb 1-\bb{k\eps_0/\eps+\a-\sqrt{k^2+1}\over
k\eps_0/\eps+\a+\sqrt{k^2+1}}\eb^2\exp(-2l\sqrt{k^2+1})\eb\nn\\
P_d&=&~2kT\mu\bara^2{d \over dl}F_0(l) + kTm^3{d \over
dl}F_{(vdW)}(l),\label{lin_disj} \eea 
where $l=Lm,~\a = m\g,\bara =m\barg$ with $\g = \barg = \mu^*/2\mu$. 
Typical values for these
parameters are 
\be kT = 4.~10^{-21}\;J,~\mu \sim 0.2\;mM \Rightarrow m
\sim 0.05\;nm^{-1},~\a = \bara = 2.0\;.  \ee 
The coefficients in Eq. (\ref{lin_disj}) are then 
\be
c_0~=~2k_BT\mu\bara^2~=~4000\;Pa,~~c_1~=~k_BTm^3~=~500\;Pa,~~\e/\e_0~=~80\;,
\ee 
where the ratio $\e/\e_0$ is for water to air.  The disjoining
pressure $P_d$ is given in Pascals. For these values a plot of $P_d$
is shown in Fig. \ref{Pd}. The solid curve is the total value of $P_d$
while the dashed curve gives the repulsive contribution from $F_0$ and
the attractive contribution from $F_{(vdW)}$ is shown as the dotted
curve. The theory predicts a collapse certainly by $L = 8\;nm$ and
since the collapse corresponds to a first order phase transition the
Maxwell construction will predict that collapse will be observed at
larger $L$. Since we do not have a theory for the short-range
repulsive force that eventually stabilizes the NBF we cannot use the
Maxwell construction to give an accurate value for where the film
becomes metastable but $8\;nm < L < 20\;nm$ would be a reasonable range.

The attractive van der Waals contribution $P_d^{(vdW)}$ above can be
compared with the Hamaker form at short distances
predicted by Eq. (\ref{pvdwasym}). However, numerical study of the case of 
interest here shows that the values of $L$ for which this
behavior holds are too small to be relevant to the collapse described
above.

For $L \gg 1/m$ the leading term is given by Eq. (\ref{hamasy})
and the value of $\a$ is crucial in determining the overall
coefficient. Indeed, for $\a \sim 1$ the behavior will given by
non-leading terms not shown here. In Fig. \ref{pvdw} we plot
$P_d^{(vdW)}$ versus $L$. It is seen that while the large $L$
asymptotic form (\ref{hamasy}) is a good approximation 
for $L > 1/m = 20nm$, in the the region important to the collapse, 
$5nm < L < 20nm$, the full
result deviates strongly from this form. Thus we find that the full
expression for $P_d^{(vdW)}$ must be used in the region of interest.

The surface-charging mechanism is very important to the prediction of
the collapse transition.  If a fixed surface charge is used we should
omit the quadratic term in $\phi$ in the expansion of the source in
Eqs. (\ref{limit},\ref{scharge}). This corresponds to setting $\a=0$
where is occurs explicitly in the expressions for $F_0$ and
$F_{(vdW)}$ in Eq. (\ref{lin_disj}) while not changing the values of
the coefficients $c_0, c_1$. In this case for the parameters above
there is no collapse.

In a more general case with both $\mu_+^*,\mu_-^*$ non-zero the same
expressions as in Eq. (\ref{lin_disj}) apply but with the
generalization $\a \neq \bara$.  The effect of choosing fixed $\bara =
2.0$ but varying $\a$ can be seen in Fig. \ref{alpha} where the values
used above of $c_0 = 4000~Pa,~~c_1 = 500~Pa$ are adopted but different
values of $\a = 1.0,1.5,2.0,3.0$ are used. The term most affected is
the repulsive mean-field term and even in the linearized theory this
is very sensitive, as we should expect, to the charging mechanism for
the surfaces. We see that the predicted properties of the collapse
transition are strongly dependent on the choice of $\bar{\a}$ and $\a$
and therefore on the details of the film being studied.

\section{Behavior of the Surface Charge}
In the limit $\delta \to 0$ the surface charge (on one surface) per
unit area, $\sigma$, can be seen to be 
\ben \sigma = -{e\over 2}\left(
\mu_+^*{\partial \over \partial\mu_+^*} - \mu_-^*{\partial \over
\partial\mu_-^*}\right) \beta J ~=~-{e\over 2}\bb \g{\partial  \over \
\partial\barg}
+ \barg{\partial \over \partial \g}\eb \gb J\;, 
\een 
where $\g,~\barg$ are defined in Eq. (\ref{gammas}).

Even in the Gaussian approach the source terms encode the non-trivial
charging properties of the surfaces bounding the ionic liquid. From
Eq. (\ref{limit}) we have used the approximation for the source 
\ben
\mu^*_+\exp(i\gb e\phi)+\mu^*_-\exp(-i\gb e\phi)~=~ i\l\phi~-~\half\gb
\eps m^2\g\phi^2~+~\ldots~.\label{scharge} 
\een 
The term linear in $\phi$ represents a fixed surface charge but the term 
in $\phi^2$ corresponds to surface charge fluctuations.

We decompose $\sigma$ in terms of the mean field contribution and the
van der Waals contribution $\sigma = \sigma_0 + \sigma_{vdW}$.  The
term coming from the mean field (and constant or ideal) contributions to the
grand potential is 
\ben \s_0(L) = {e \mu \over m} {\bara \over (\a +
\tanh({mL\over 2}))} \left[ 2 \tanh({mL\over 2}) + {\bara^2 \over (\a
+ \tanh({mL\over 2}))} \right].  
\een 
The contribution from the van der Waals term is independent of 
$\barg$ and can be written as 
\ben
\s_{vdW} = -{e\over 2} \bara {\partial \over \partial \a} \beta
J_{vdW}.  
\een 
We find 
\ben \s_{vdW} = {\bara e m^2\over 2 \pi} \int
dk\; k\; {f(k)\over 1 - f^2(k)}\; {\sqrt{k^2 + 1}\over (k\e_0/\e + \a
+ \sqrt{k^2 + 1})^2}\; \exp(-l\sqrt{k^2 + 1}), 
\een 
where $f$ is given
by 
\ben 
f(k) = \left({k\e_0/\e + \a -\sqrt{k^2 + 1} \over k\e_0/\e +
\a + \sqrt{k^2 + 1}} \right) \exp(-l\sqrt{k^2 + 1}).  
\een 
For the parameters given in the previous 
section: $\mu = 0.2\; mM,\; m =
0.05\; nm^{-1},\; \a = \bara = 2.0$ the behavior of $\s$ is shown
versus $L$ in Fig. \ref{chplot}. One notices that the surface charge is
regularized on varying $L$, though not drastically 
(about $10 \%$ over $40\; nm$) just up to  the film
thickness $L \sim 10 {\rm nm}$. One sees that in this case
the effect of the van der Waals term is to decrease the value of $\sigma$
from its mean field value. 

\section{Conclusions and Outlook}
In this paper we have presented a field theoretic formulation of the 
electrostatic interactions in soap film like systems, which treats on
the same footing the zero frequency van der Waals or Lifshitz 
terms  and the contributions coming from ionic fluctuations.
The basic idea  is to use the static part of the QED Lagrangian
coupled to the charge density coming from the ions in the system
and then integrating over the electrostatic potential $\psi$ and 
the positions of the ions (which are treated classically). The 
time dependent and magnetic field terms in the full QED Lagrangian are
thus neglected, this is equivalent to the non retarded limit where 
the velocity of light $c \to \infty$. Retardation effects can be taken
into account by summing over the non zero Matsubara frequencies, however
the coupling of these terms with the ionic distribution is weak. The 
incorporation of retardation effects requires in addition the frequency 
dependence of the electric permittivies.   

This 
treatment is easily applicable to systems with spatially varying dielectric 
constants and elegantly avoids calculations of the arising image effects.
The formalism also allows the incorporation of surface charges induced 
by equilibrium processes.   
In the grand canonical ensemble one obtains a Sine-Gordon field theory.
Linearizing this theory leads to a soluble Gaussian field theory 
and is equivalent to the Debye-H\"uckel approximation, which should be 
valid for weak ionic concentrations. In this form the evaluation
of the grand potential is carried out by using the Feynman kernel for
simple harmonic oscillators. The use of the Feynman kernel in the 
field theoretic formalism allows us reproduce 
a wide range of results established in the literature via other methods. 
The effect of surface charge fluctuations are 
can also be taken into account and it was also shown 
how many different layers of varying dielectric constant simply lead to a 
renormalization of the simple bilayer result.

Preliminary investigation of the relevant experimental soap film parameters
shows that the van der Waals contributions lead to a weak (screened)
attraction at large intersurface separations. For thinner films there
is an increased attraction, which can overcome the mean field repulsion
present in the models considered here. However in the region where the 
film collapses, the attraction does not have the simple Hamaker $1/L^3$
form and is strongly dependent on the Debye mass $m$ and the surface
surface-charging parameters. Indeed an essential
ingredient is the inclusion of the surface-charging mechanism which,
although treated here in the Gaussian approximation, nevertheless
predicts an $L$-dependent surface charge density which is important to
the details of the collapse. A linear approximation to the surface
charge source is inadequate since it leads to a fixed surface charge
and hence a diverging mean-field repulsive pressure $P_0$ as $L$
decreases. By including the term in $\phi^2$ in the expansion of the
sources, the surface charge is shown to decrease as $L$ decreases and
so the divergence in the mean-field repulsion $P_0$ is regulated.

Although not necessary for the theory presented here it is interesting to
compare the outcome with our work on the 1D coulomb gas model for a
soap film \cite{dehose}. The mechanism in 1D for the collapse was the
changing balance of contributions to $P_d$ between the even and odd
eigenfunctions of the Mathieu equation as the film thickness $L$
varied. The important states were the lowest lying ones including the
ground-state. In the Gaussian approximation used in the present work
we can ask which are the important eigenfunctionals of the theory in
3D which play a similar role. In this model the eigenfunctionals are
products of harmonic oscillator eigenfunctions for each of the
transverse momentum modes separately for which the coordinate is
$\tilde{\phi}(\bp)$. The source term is, however, only a function of
the zero mode $\tilde{\phi}(0)$ and the important corresponding term
in the wavefunctional from Eq. (\ref{hamiltonian}) is ($\bp=0,~X
\equiv \tilde{\phi}(0)$) 
\be \Psi_n(X)~=~h_n(X)\exp(-mX^2/2)\;, 
\ee
where $n$ is the oscillator excitation number. For large $n$ the
Hermite polynomial has an oscillatory factor 
\be
~h_n(X)~\sim~\cos(\sqrt{n}\;mX-(n-1)\pi/2)\;.  
\ee The relevant
coefficient is the overlap of this wavefunction with the source term
in Eq. (\ref{P=0}) $\exp(i\sqrt{A}\mu^*\gb eX)$ and it is clear that
this will not be large unless the oscillatory factors match. This will
be the case when 
\be 
\sqrt{n}\;m~\sim~\sqrt{A}\mu^*\gb e~, 
\ee 
and the
corresponding energy values are of order $E \sim (n+1/2)m \sim
A{\mu^*}^2(\gb e)^2/m$ giving an extensive contribution to the free
energy as we must expect. Hence the important states are not the
ground-state and those nearby but highly excited states which carry
the extensive nature of the system.

By taking experimentally reasonable values for the parameters in our
formulae we obtain acceptable predictions for the collapse phenomenon
and surface-charging for simple description of the film. However, to
make accurate predictions will require the film to be modeled as a
multiple layer with the correct permittivities for each layer and
possible charging potentials included. We must also include the
contributions of non-linear and non-Gaussian operators and to do this
involves three ingredients. The first is to solve the non-linear
mean-field equations, the second is to develop the perturbation theory
for the non-Gaussian source operators within the Gaussian field theory
and the third to use perturbation theory for the non-Gaussian
interactions given by the Sine-Gordon theory in the film interior. Because
the system is not translation invariant these perturbation theories is
not standard but it has been developed and will be presented in a
succeeding paper together with the non-linear mean-field formulae.

{\bf Acknowledgment:} We would like to thank D. Bonn and I.T. Drummond for
interesting discussions.

\befh
\bec
\epsfig{file=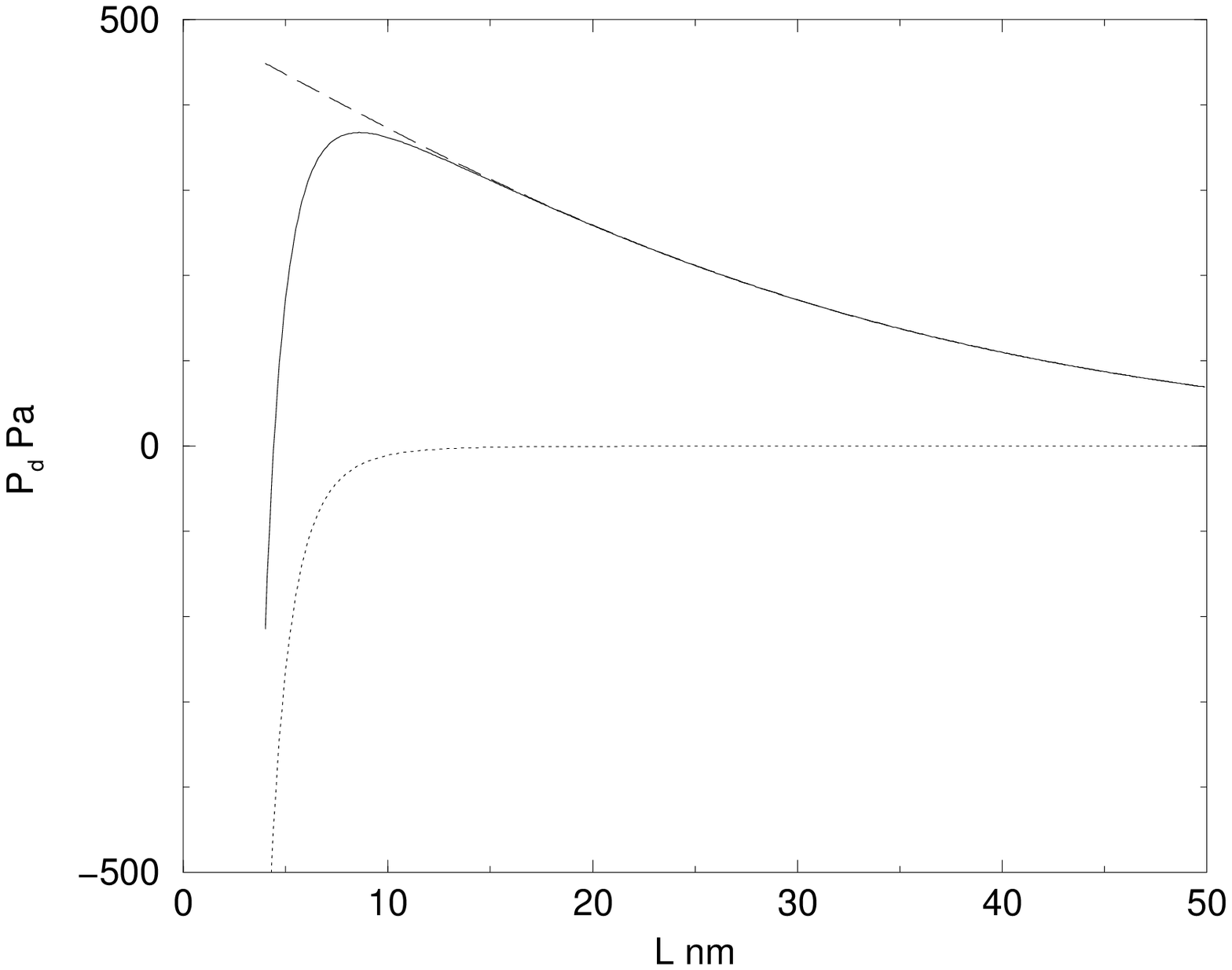,height=80mm}
\enc
\caption{\label{Pd}\small
The disjoining pressure $P_d$ for the linearized theory using typical parameter values:
$\mu_{\rm NaCl}=0.2\;mM,~m_{\rm Debye}=0.05\;nm^{-1},~\a = 2.0$.
The solid line is the full result, the dashed line is the repulsive contribution
and the dotted line the attractive, van der Waals contribution.
}
\enf
\befh
\bec
\epsfig{file=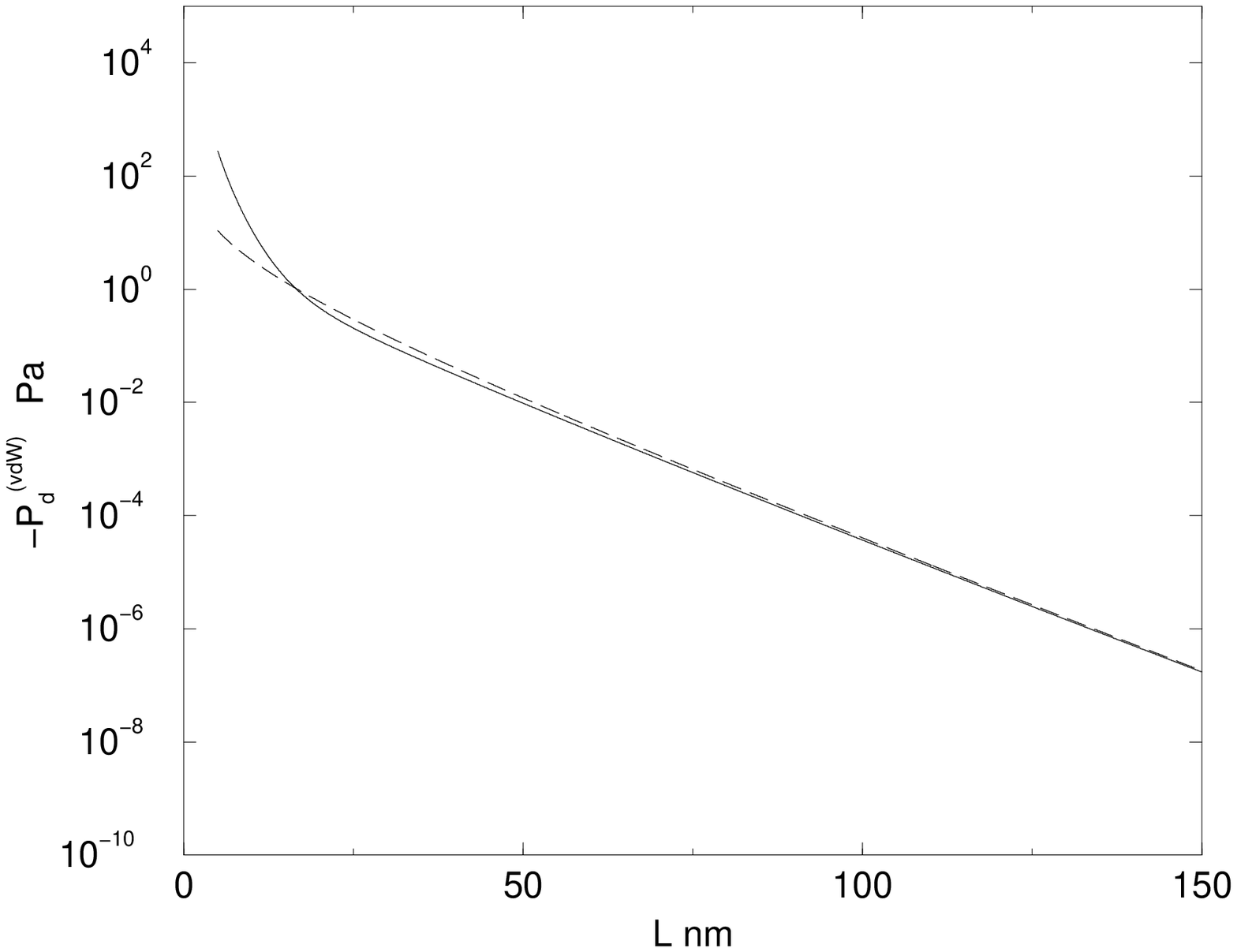,height=80mm}
\enc
\caption{\label{pvdw}\small
The van der Waals contribution $P_d^{(vdW)}$ for the linearized theory plotted against $L$
using typical parameter values: $\mu_{\rm NaCl}=0.2\;mM,~m_{\rm Debye}=0.05\;nm^{-1},~\a = 2.0$.
The dashed line is the asymptotic formula Eq. (\ref{hamasy}) which is seen to be accurate until
$L \leq 1/m = 20 nm$. The standard Hamaker form $\propto L^{-3}$ is not applicable for the
relevant values of $L$.
}
\enf
\befh
\bec
\epsfig{file=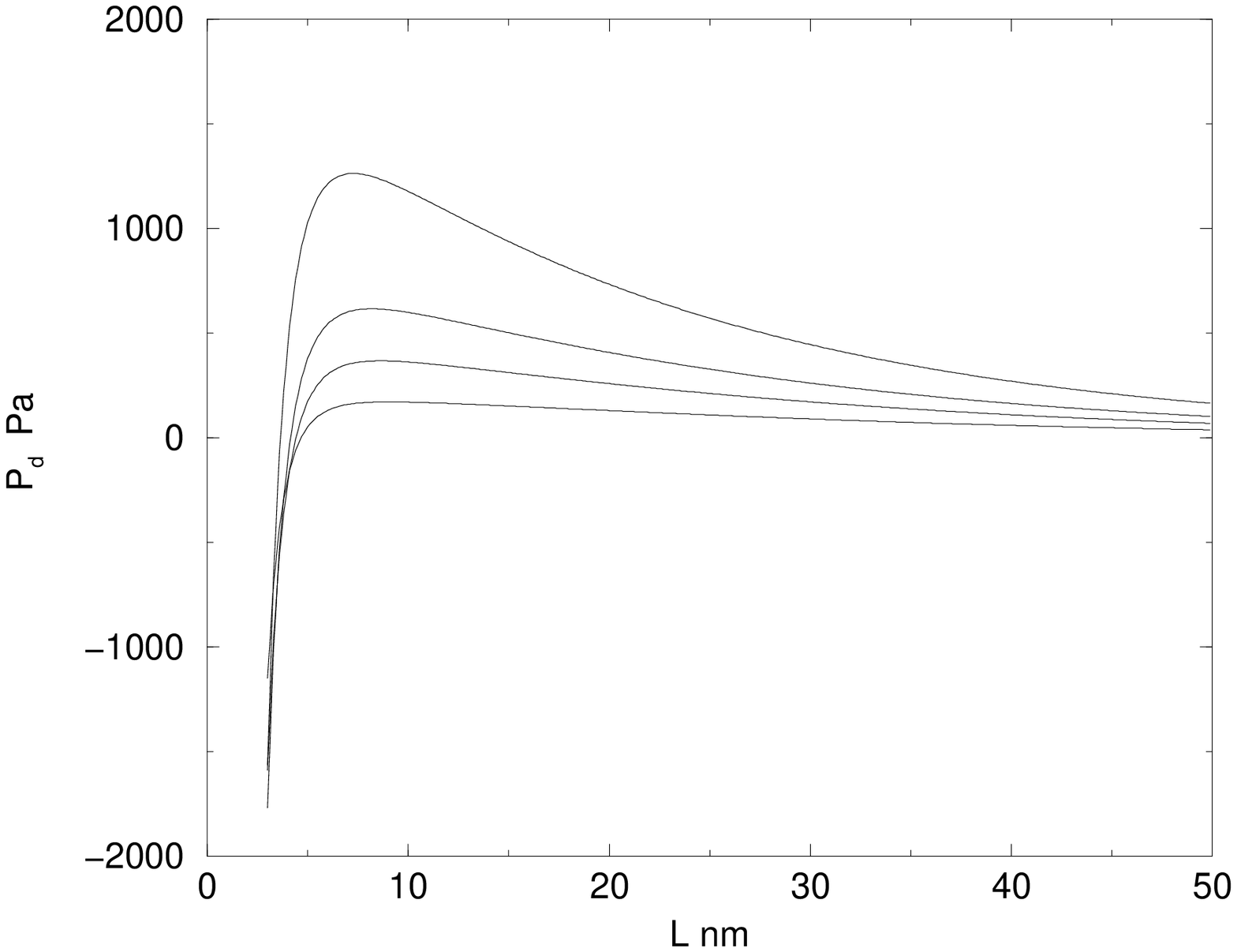,height=80mm}
\enc
\caption{\label{alpha}\small
The disjoining pressure $P_d$ for the linearized theory using typical parameter values:
$\mu_{\rm NaCl}=0.2\;mM,~m_{\rm Debye}=0.05\;nm^{-1},~\bara = 2.0$
but for different values of $\a$ occurring explicitly in Eq. (\ref{lin_disj}).
The curves from top down are for $\a = 0.5, 1.0, 1.5, 2.0, 3.0$. 
}
\enf

\befh
\bec
\epsfig{file=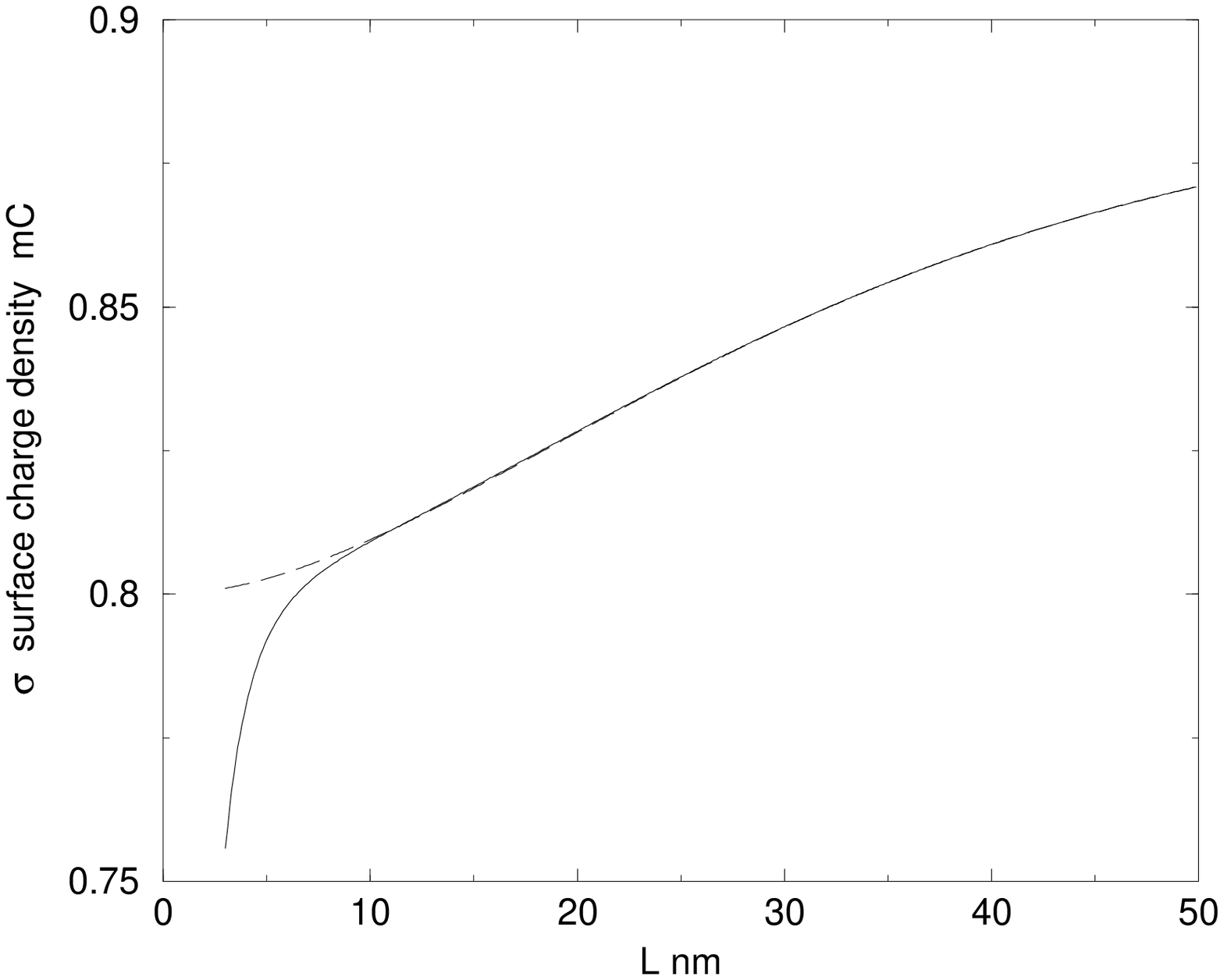,height=80mm}
\enc
\caption{\label{chplot}\small
The surface charge $\s$ in $mC$ plotted versus $L$ for
$\mu_{\rm NaCl}=0.2\;mM,~m_{\rm Debye}=0.05\;nm^{-1},~\a = 2.0$.
The solid line is the full result, the dashed line is the mean field contribution $\s_0$.
The van der Waals contribution, $\s_{(vdW)}$ is negligible until $L \leq 10 nm$
where it is a responsible for the rapid decrease in $\s$. 
}
\enf

\baselineskip =18pt
\begin{thebibliography}{0}
\bibitem{is}{J. Israelachvili, {\em Intermolecular and Surface
forces},
(Academic Press) (1992)}
\bibitem{rusasc}{W.B. Russel, D.A. Saville and W.R. Schowalter,
{\em Colloidal Dispersions} (Cambridge University Press, Cambridge) (1989)}
\bibitem{my}{K. Mysels and M.N. Jones, Discuss. Faraday Soc. {\bf 42}, 42 
(1966)}
\bibitem{exkokh}{D. Exerowa, D. Kolarov and Khr. Khristov, Colloids Surfs.
{\bf 22}, 171 (1987)}
\bibitem{dese}
{D.S. Dean and D. Sentenac, Europhys. Lett. {\bf 38}, 9, 645, (1997);
D. Sentenac and D.S. Dean J. Colloid Interface Sci. {\bf 196},
35 (1997)}
\bibitem{pisa}{P.A. Pincus and S.A. Safran, Europhys. Lett. {\bf 42}, 103
(1998)}
\bibitem{caetal}{V. Castelleto et al preprint ENS (2001)}
\bibitem{nipa1}{Ninham B.W. and Parsegian A., J. Theor. Biol 
{\bf 31}, 405, 1971}
\bibitem{orsteric}{I. Borukhov, D. Andelman and H. Orland, Phys. Rev. Lett. 
{\bf 79}, 435 (1997)}
\bibitem{bebe}{O. B\'elorgey and J.J. Benattar, Phys. Rev. Lett. {\bf
66}, 313 (1991)}
\bibitem{po}{R. Podgornik, J. Chem. Phys. {\bf 91}, 5840,(1989)}
\bibitem{codu}{R.D. Coalson and A. Duncan, 
J. Chem. Phys. {\bf 97}, 5653, (1992)}
\bibitem{btco}{N. Ben-Tai and R.D. Coalson, J. Chem. Phys. {\bf 101}, 5148, (1994)}
\bibitem{lif}{I.E. Dzyaloshinskii, E.M. Lifshitz and L.P. Pitaevskii,
Advan. Phys. {\bf 10}, 165, 1961}
\bibitem{mani}{J. Mahanty and B. Ninham, {\em Dispersion Forces}
(Academic, London), (1976)}
\bibitem{podo}{R. Podgornik and J. Dobnikar, cond-mat 0101420, (2001)}
\bibitem{kago}{M. Kardar and R. Golestanian, Rev. Mod. Phys. {\bf 71}, 1233 
(1999)}
\bibitem{motr}{V.M. Mostepanenko and N.N. Trunov, {\em The Casimir Effect and
its Applications} (Clarendon Press, Oxford) (1997)}
\bibitem{dehose}{D.S. Dean, R.R. Horgan and D. Sentenac, J. Stat. Phys. 
{\bf 90}, 899 (1998)} 
\bibitem{edle}{S. Edwards and A. Lenard, J. Math. Phys. {\bf 3}, 778,
(1962)}
\bibitem{teme}{G. T\'ellez and L. Merch\'an cond-mat 0109377 (2001)}
\bibitem{van}{N.G. Van Kampen, B.R.A. Nijboer and K. Schram Phys. Letters {\bf 26A}, 307 (1968)}
\bibitem{atmini1}{P. Attard, D.J. Mitchell and B.W. Ninham,
J. Chem. Phys. {\bf 88}, 4987 (1988)}
\bibitem{atmini2}{P. Attard, D.J. Mitchell and B.W. Ninham,
J. Chem. Phys. {\bf 89}, 4358 (1988)}
\bibitem{atkjmi}{P. Attard, R. Kjellander and D.J. Mitchell and B. J\"onsson 
J. Chem. Phys. {\bf 89}, 1664 (1988)} 
\bibitem{sebe}{D. Sentenac and J.J. Benattar, Phys. Rev. Lett. {\bf 81}, 160
(1999)}
\bibitem{nipa2}{B.W. Ninham and V.A. Parsegian, J. Chem. Phys. {\bf 52}, 4578 (1970)}
\bibitem{mepist}{R. Menes, P. Pincus and B. Stein, 
Phys. Rev. E. {\bf 62}, 2981 
(2000)}
\bibitem{dorivr}{W.A.B. Donners, J.B. Rijnbout and A. Vrij,
J. Colloid Interface Sci. {\bf 60}, 540 (1997)}
\bibitem{fehi}{R. Feynman and A.R.  Hibbs, {\em Quantum Mechanics
and Path Integrals} (McGraw-Hill) (1965)}
\bibitem{balo}{D. Bailin and A. Love, {\em Introduction to Gauge Field Theory}
(Adam Hilger, Bristol and Boston) (1986)}
\end{thebibliography}
\end{document}